\documentclass[times,twocolumn,final]{elsarticle}

\usepackage{cag}
\usepackage{framed,multirow}
\usepackage{booktabs}
\usepackage{multirow}
\usepackage{chngcntr}
\counterwithout{table}{section}
\usepackage{hyperref}

\newcommand{\revise}[1]{\textcolor{black}{#1}}

\usepackage{amssymb}
\usepackage{amsmath}

 \usepackage{lineno}

\journal{Computers \& Graphics}

\begin{document}

\begin{frontmatter}



\title{A Unified Framework for Interactive Visual Graph Matching via Attribute-Structure Synchronization}

\author{%
  Yuhua Liu\textsuperscript{a}, 
  Haoxuan Wang\textsuperscript{a}, 
  Jiajia Kou\textsuperscript{a}, 
  Ling Sun\textsuperscript{a}, 
  Heyu Wang\textsuperscript{a,*}, 
  Yongheng Wang\textsuperscript{b}, 
  Yigang Wang\textsuperscript{a}, \\
  Jinchang Li\textsuperscript{c} and
   Zhiguang Zhou\textsuperscript{a,*}%
}

\address{
  \textsuperscript{a} Hangzhou Dianzi University, Hangzhou 310018, China\\
  \textsuperscript{b} Zhejiang Lab, Hangzhou 311121, China\\
  \textsuperscript{c} School of Data Sciences, Zhejiang University of Finance \& Economics, Hangzhou 310018, China
}

\begin{abstract}
In traditional graph retrieval tools, graph matching is commonly used to retrieve desired
graphs from extensive graph datasets according to their structural similarities. However, in real applications, graph nodes have numerous attributes which also contain valuable information for evaluating similarities between graphs. Thus, to achieve superior graph matching results, it is crucial for graph retrieval tools to make full use of the attribute information in addition to structural information. We propose a novel framework for interactive visual graph matching. In the proposed framework, an attribute-structure synchronization method is developed for representing structural and attribute features in a unified embedding space based on Canonical Correlation Analysis (CCA). To support fast and interactive matching, \revise{our method} provides users with intuitive visual query interfaces for traversing, filtering and searching for the target graph in the embedding space conveniently. With the designed interfaces, the users can also specify a new target graph with desired structural and semantic features. Besides, evaluation views are designed for easy validation and interpretation of the matching results. Case studies and quantitative comparisons on real-world datasets have demonstrated the superiorities of our proposed framework in graph matching and large graph exploration.
\end{abstract}





%



\end{frontmatter}

\makeatletter
\let\@fnsymbol\@empty  
\makeatother

\renewcommand\thefootnote{}  

\footnotetext{%
  \parbox{0.9\columnwidth}{
    \textsuperscript{*}Corresponding author. \\
    Yuhua Liu and Haoxuan wang contributed equally to this work.
    
    Email addresses: 
    liuyuhua@hdu.edu.cn (Y. Liu),
    
    whxxyhf111@gmail.com (H. Wang), 
    
    241330014@hdu.edu.cn (J. Kou), 
    
1055389464@qq.com (L. Sun),

heyuwang@hdu.edu.cn (H. Wang), 

wangyh@zhejianglab.org (Y. Wang), 

    yigang.wanghdu.edu.cn (Y. Wang),
    
    ljc@zufe.edu.cn (J. Li),
    
    zhgzhou@hdu.edu.cn (Z. Zhou).
  }%
}
\renewcommand\thefootnote{\arabic{footnote}}  



\section{Introduction}\label{sec1}

The advancement of graph data management techniques has led to the widespread collection of numerous graph datasets, catering to the data-dependent investigations and applications across various domains, such as social connections \cite{liu2022protect}, business relations \cite{henna2021enterprise}, and knowledge derivations \cite{ji2021survey}. Graph matching frequently serves as an exploratory method for retrieving desired complete graphs from a large-scale graph dataset with extensive graphs based on various similarity measures. 

Graph matching prioritizes structures, with their similarity measured in terms of topological features or learned vectors. For instance, in a chemical compound database $D$, biochemists often retrieve and match the desired compounds in $D$ based on the structure similarity. To achieve a more comprehensive representation of graph structures in matching tasks, techniques such as graph kernel \cite{kriege2020survey} and graph representation learning \cite{song2022graph} have been applied alongside traditional graph similarity metrics, including maximum common subgraph (MCS) \cite{10496270} and minimum graph edit distance (MGED) \cite{9212900}. 

However, in practical scenarios, graph nodes are depicted with rich attributes, which are also important for graph matching \cite{mahmood2023interactive}. For instance, in the analysis of protein-protein interaction networks, incorporating protein attributes enables biologists to achieve more precise identification of protein complexes \cite{zhou2022heterogeneous}. Similarly, within a large-scale air propagation network, the air transmission structure and the proportion of urban air composition constitute important propagation patterns \cite{deng2019airvis}.

Evidently, the integration of both structural and attribute information into a unified model provides substantial advantages for reliable graph matching. Nevertheless, it is challenged from three aspects as follows:

\textbf{C1}: Structures and attributes exhibit distinct and heterogeneous features. Constructing an attribute-structure synchronization considering the correlation between them toward yielding better-quality matching results is challenging.

\textbf{C2}: Graph query language serves as a prevalent means of retrieving graphs. However, it is difficult to learn and use for desired graph matching, especially for those non-experts. Providing visual graph querying interfaces is a feasible manner to synchronize structures and attributes, and empower non-experts to construct queries. 

\textbf{C3}: Graph matching, employing attribute-structure synchronization, often retrieves numerous candidate graphs. It becomes essential to furnish users with quantitative and visually accessible recommendations, aiding them in selecting the most suitable graph to meet their specific requirements.

To tackle the above challenges, we propose a method to interactively query graphs considering attribute-structure synchronization. For a specific graph database, our method will train a model to learn the potential correlation between attribute and structure. There are two stages: 1) the representations of structure features and attributes of graphs and 2) the fusion considering the intrinsic correlation between two dimensional features. Graph representation learning is commonly used to learn the structure feature of graphs. In the first stage, Graph2vec is employed to learn the structure feature and the multiple attributes of nodes are statistics as multi-dimensional vectors. Then Canonical Correlation Analysis (CCA) is utilized to integrate the acquired structural information and multiple attributes into a synthetic model. This model is then subjected to joint dimensionality reduction, resulting in the projection of graphs into a low-dimensional feature space, where those graphs with shared structural and attribute characteristics are positioned close to each other \textbf{(C1)}. After obtaining the joint representation of the graph database, our method facilitates the measurement of distance-based similarity, enabling efficient graph matching.

To further facilitate graph matching with attribute-structure synchronization, we have devised two visual matching schemes. The one scheme provides multiple projection views to assist users in quickly traversing, filtering and searching for the target graph during the matching process. The other provides a variety of customizable panels, empowering users to define a new target graph with specific semantic and structural features, and then retrieve those graphs matching the manually-specified target \textbf{(C2)}. Multiple coordinated views and a series of interactions have been designed to support the visual evaluation and exploration of matched results. Users can readily assess the similarities between the target graph and the matched results using node-link diagram and parallel coordinates \textbf{(C3)}. Finally, a visual analysis system is developed to integrate the above matching model, visualization views and interactions, enabling users to accurately match and explore large-scale multivariate graph data, as shown in Figure ~\ref{fig1}. The primary contributions of this paper are summarized as follows: 

• An attribute-structure synchronization based on Canonical Correlation Analysis is proposed for credible graph matching. 

• Visual interfaces and user-friendly interactions are designed to replace conventional graph query languages for graph matching, evaluation, and exploration. 

• Case studies and quantitative comparisons based on real-world datasets are conducted to verify the effectiveness and convenience of our system.

\section{Related work}\label{sec2}

\subsection{Graph Matching}
\label{subsec2}

Graph matching can be divided into two categories, such as decision version and matching version. The decision version is to determine whether the query graph exists in a graph database. The matching version refers to discovering the embedding of the query graph in a large graph. Previous matching methods are categorized as filter-then-verify (FTV) \cite{wangmo2021subtempora} and direct subgraph isomorphism (SI) \cite{bi2016efficient}. The FTV methods include two stages: filtering and verification. In the filtering stage, the query graph is decomposed into features such as paths \cite{cheng2024l2match}, trees \cite{karpov2022syncsignature}, or their combinations. The graphs without these features are filtered out, and the remaining graphs are retrieved to form a candidate set. In the verification stage, the final querying results are generated by conducting a subgraph isomorphism test to the query graph against the candidate set. However, these technologies suffer from limited filtering capabilities and high cost of verification \cite{chai2024asm}. To address this issue, the direct SI method \cite{wang2024pathlad+}, was proposed, which accelerates the SI test by locating the best candidate nodes for each graph, and joins these nodes to match the query \cite{cheng2023l2match}. The proposed FTV and direct SI are both applied to address the two versions of graph matching problems. 

In the field of visualization, graph matching belongs to the matching version, i.e., exploring the embedding of a target structure/subgraph from a large network \cite{yang2023survey,lyu2024visualization}. For example, VISAGE \cite{pienta2016visage} performs graph querying on a large graph with nodes of different types and attributes. g-Miner \cite{cao2015g} supports mining groups from a multivariate network with the capability to adapt to complex or dynamic requirements of the desired group. Wong et al. \cite{wong2015visual} developed a visual analytics paradigm, which cannot only rapidly preprocess a graph with billions of edges but also support graph query and interactive visualization on the processed graph. Chen et al. \cite{chen2018structure} developed a visual exploration system that suggests appropriate structures upon user-specified exemplars in a large network. VIGOR \cite{pienta2017vigor} is an exemplar-based interaction technique to support top-down and bottom-up searches for paradigm patterns from a large network. Song et al. \cite{song2021interactive} use graph representation learning to match graphs with similar subgraph structures. \revise{However, most existing works treat structure and attribute either independently or fuse them through predefined or tightly coupled models, limiting generalizability and interpretability. our approach introduces a modular and extensible framework that decouples structure embedding and attribute alignment via Canonical Correlation Analysis (CCA), enabling flexible integration of heterogeneous graph data. It uniquely combines embedding, synchronization, and visual analytics to support interpretable and interactive graph-level matching.}

\subsection{Graph Representation Learning}

Graph representation learning (GRL) focuses on how to embed graph structures into a vector space while preserving the structural property, which can be roughly divided into three categories \cite{zhou2020context,zhou2024chartkg}. (1) Factorization based GRL methods represent the connection between nodes in the form of an adjacency matrix, Laplace matrix \cite{zhou2022imgc}, etc., and then factorize these matrices to obtain node vectors. (2) Random-walk based GRL methods generate node sequences by a random walk on the graph and obtain the representation results so that nodes have similar embeddings if they tend to co-occur on random walks, such as DeepWalk and node2vec \cite{grohe2020word2vec}. Inspired by the random walk strategy, Graph2vec is proposed to learn representations of whole graphs by characterizing each graph as a sequence of rooted subgraphs. (3) Deep-learning based GRL methods utilize the ability of deep auto-encoder to generate vector representations, which can capture the nonlinear structure information in graphs \cite{tang2022graph}. 

Attribute graph embedding has attracted much attention in recent years, since it’s strongly believed that graph semantics (e.g., node attributes) carries complementary knowledge beyond the topology of graphs. Two categories of methods are proposed. (1) The matrix-decomposition based methods \cite{yang2023pane} are to add the attribute information of nodes to the process of matrix decomposition, for further generating new node vectors. (2) The deep-learning based method \cite{wang2024dgnn} projects the attribute information and topology information of nodes into the same semantic space. However, the above methods focus on the embedding of nodes in attribute graphs, which is different from our study which applies global attribute graph embeddings to graph matching.

\subsection{Graph Similarity Measure}

It is significant for graph matching to define a feasible similarity metric \cite{wang2022dl4scivis}which is measured with the difference between two graphs by means of quantitative similarity values. GQL is a graph query language, which cannot only support filtering nodes with specific attribute values, but also defines semantic rules for graph matching, such as Cypher \cite{francis2018cypher}, Gremlin \cite{10.1007/978-3-319-64468-4_6} and SPARQL \cite{hogan2020sparql}. In addition, some graph similarity studies based on semantic features have been proposed. For example, Khan et al. \cite{khan2013nema} proposed a subgraph matching cost metric which unifies both structure and node label similarity. Zhao et al. \cite{zhao2010graph} took the local features of nodes as basic graph indexing units. 

Structure-based similarity measures are widely investigated in the field of graph databases, which can be divided into two categories depending on whether the space is transformed in the measurement process: direct methods and indirect methods. Direct methods measure the structural similarity through a graph matching procedure. The most widely-used direct methods are maximum common subgraph (MCS) \cite{10496270} and minimum graph edit distance (MGED) \cite{9212900}. By contrast, indirect methods \cite{doan2021interpretable} transform graphs into vectors, and then estimate the similarity between graphs in the vector space. Currently, graph kernels and graph representation learning have also been employed for measuring graph similarities. Graph kernel measures the similarity between a pair of graphs based on the recursively decomposed substructures of graphs \cite{kwon2017would}. Graph representation learning \cite{song2021interactive} uses the node-level or graph-level vectorized representation of graphs to measure their similarity. With the development of deep learning technology, graph neural network has also been successfully applied to learn the similarity between graphs \cite{li2022structure}.

\section{Requirement analysis and system overview}
\label{sec3}

\begin{figure}[h]
\centering
\includegraphics[width=\linewidth]{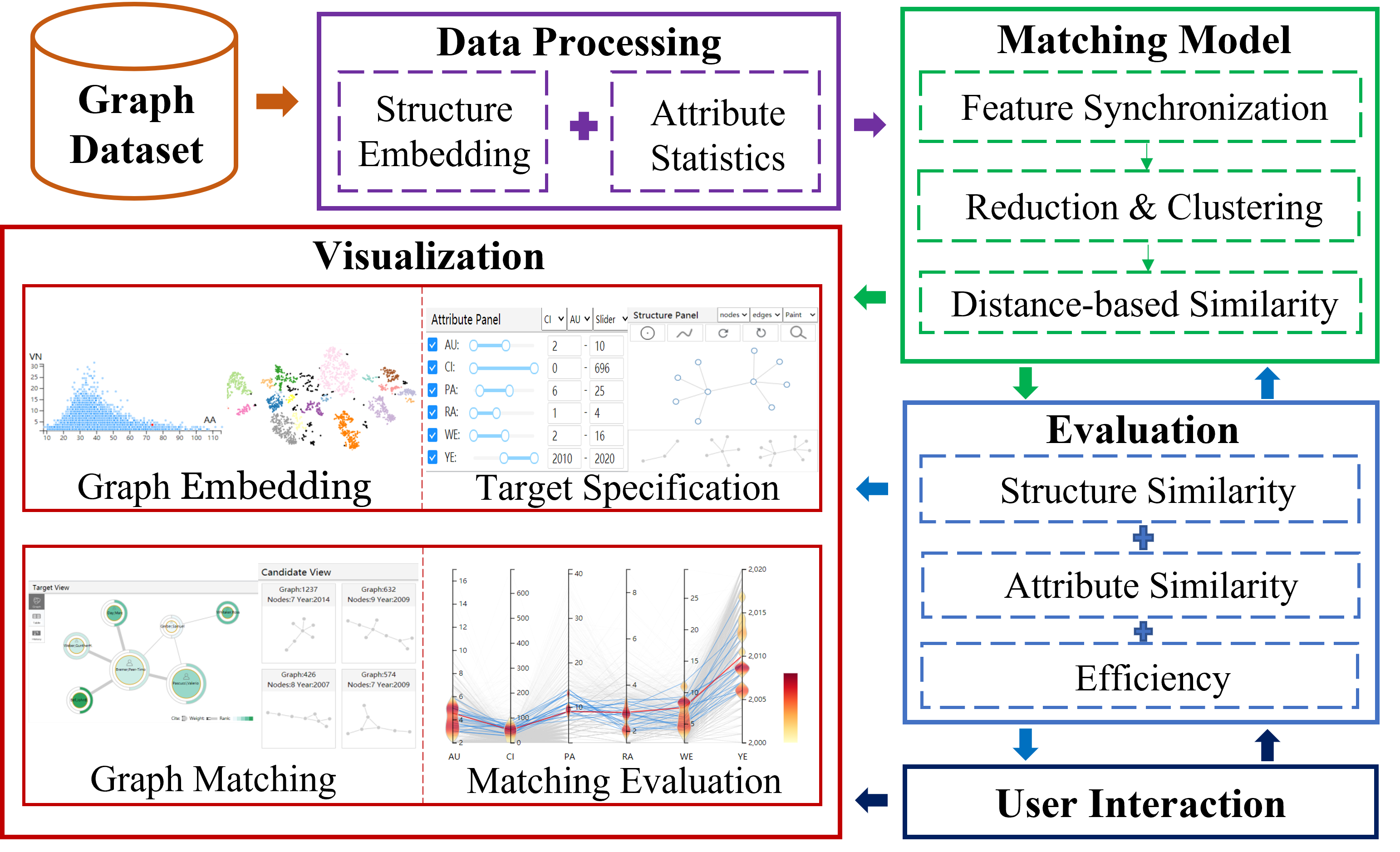}
\caption{\revise{The pipeline of our graph matching system based on attribute-structure synchronization. Users first import a dataset and configure the matching model via the control panel. A target graph can be selected from the projection view or constructed manually. The selected graph is displayed in the target view for detailed inspection. Once matching is triggered, candidate graphs with similar structure and attributes are retrieved and presented for comparison.}}\label{fig2}
\end{figure}

\subsection{Requirement Analysis}
In this study, we collaborate with three domain experts ($E_1$, $E_2$ and $E_3$)to identify key challenges in graph matching. $E_1$ specializes in graph visualization, $E_2$ is a senior data analyst focusing on social networks, and $E_3$ is a social science professor experienced in genealogical data. The design process involves two phases of expert interviews. In Phase I, experts highlighted that combining structural and attribute features is essential for graph matching, as both are critical for graph representation and insight discovery. Phase II focused on iteratively developing the system based on expert feedback, particularly for visual matching and result evaluation.From the interviews, we identify four core tasks:

\textbf{T1: Extracting structure features of graphs.} Traditional matching methods rely on substructures like paths or trees but struggle to incorporate attributes. Graph embedding transforms structures into vectors, enabling attribute-structure synchronization. Thus, selecting a suitable embedding method is essential for reliable graph matching.

\textbf{T2: Attribute-structure synchronization for graph matching.} Graph embedding methods transform graphs into a vector space, while the associated attribute information represents graphs in the attribute space. Quantifying and capturing the correlation between these two spaces is a complex challenge. Simply concatenating the structure feature and the attribute feature cannot be sufficiently helpful to make the fused feature effective for the matching purpose \cite{sargin2007audiovisual}.Therefore, a unified embedding space is needed to preserve both structural and attribute similarities.

\textbf{T3: Evaluation of graph matching results.} Locating graphs that closely resemble the target graph is achieved by quantifying the distance between graphs and identifying the nearest neighbors within the attribute-structure synthetic space.  Such distance-based graph matching needs to be further verified.  Matching results should be visually interpretable, and similarity metrics are needed to help users assess their validity.

\textbf{T4: Graph matching framework.} Current graph matching relies heavily on complex query languages, limiting usability for non-experts. Visual interfaces help but lack interactivity. A data-driven, user-friendly tool is needed to support exploratory matching based on structural and semantic conditions.

\subsection{System overview}

Motivated by the identified requirements, we design a visualization framework enabling users to conduct graph matching based on attribute-structure synchronization. The system pipeline is presented in Figure ~\ref{fig2}. First, a large-scale graph dataset is loaded into the visualization system, and two categories of features are calculated in advance, including structure features represented by the embedding technology, Graph2vec \textbf{(T1)}, and attribute features obtained with statistical methods. Then, a novel graph matching model is conducted to combine those two different features into a synthetical embedding space by the CCA-based feature fusion, for achieving distance-based similarity measures after joint dimensionality reduction \textbf{(T2)}. Furthermore, a set of well-structured views and interactions are designed for users to match a target graph of interest, evaluate and explore the matching results over this large-scale graph dataset \textbf{(T3, T4)}. We also conduct a quantitative experiment with specific metrics on two real-world datasets to assess the quality of the resultant matching results based on our method in terms of structure similarity and attribute similarity \textbf{(T3)}.

\section{Graph matching model}
\label{sec4}

\begin{figure}[h]
    \centering
    \includegraphics[width=\linewidth]{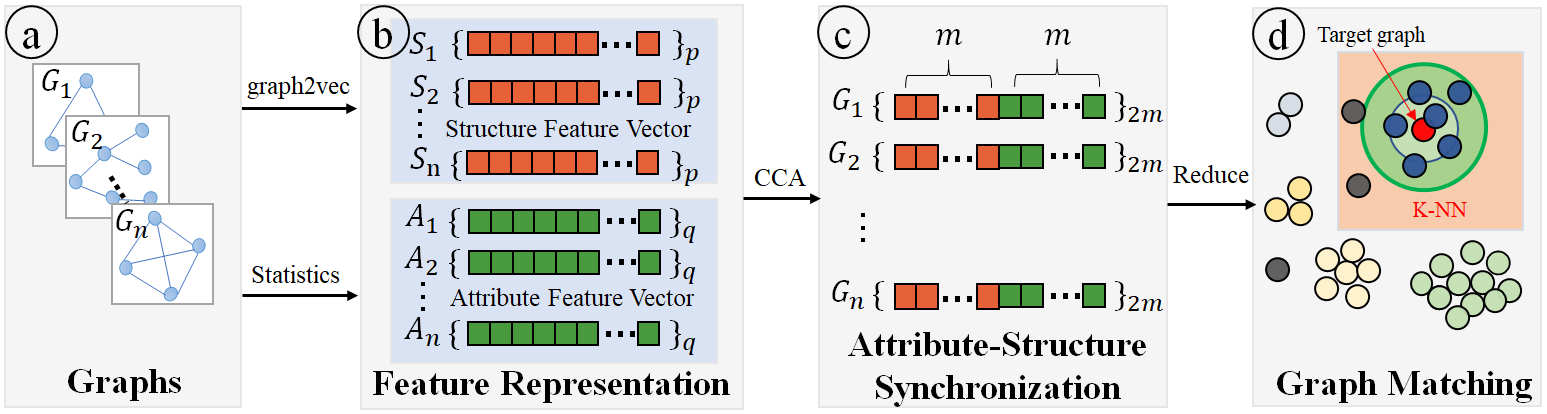}
    \caption{Given a set of n graphs (a), we extract the topological feature vectors by Graph2vec and the attribute feature vectors by a statistical method (b). Next, the two kinds of feature vectors with different lengths are translated into new uniform-length vectors and then combined into a synthetic embedding space by CCA (c). A distance-based graph matching scheme is conducted to the space for generating the matching result of a target graph (d).}
    \label{fig3}
\end{figure}

In this section, we detail our graph matching model based on the attribute-structure synchronization as shown in Figure ~\ref{fig3}.

\subsection{Structure feature extraction}

Most graph embedding technologies focus on the node-level structure instead of the global structures of graphs. For this reason, the graph-level embedding technology, Graph2vec, is introduced as our structural feature embedding model. Besides, Graph2vec is a data-driven distributed representation learning approach. Its true superiority of efficiency and robustness has been revealed when training large volumes of graphs. The Graph2vec algorithm mainly consists of two steps as follows: 

\textbf{Step1: Rooted subgraph generation.} To facilitate learning graph embedding, a rooted subgraph around each node of one graph need to be generated by such a function, GetWLSubgraph ($n$,$G_i$,$d$), which follows the WeisfeilerLehman relabeling process \cite{shervashidze2011weisfeiler}. Where n is a rooted node, $G_i$ is the graph from which the subgraph is extracted, and d is the degree of the intended subgraph. First, we obtain all neighbors of the rooted node n with the breadth-first traversal. Then we get the degree $d$ - 1 subgraph of each neighboring node until $d$ = 0. In this recursive process, we concatenate the list which is used in advance to store subgraphs for generating the $d$ -degree subgraph $sg_{n}^{\left( d\right) }$ n of n. Given the graph $G_i$, a set of rooted subgraphs around other nodes in $G_i$ can be generated in the same way, thus the context of $G_i$ is denoted as c($G_i$)= $\left\{ sg_{1}, sg_{2},\cdots \right\} $. 

\textbf{Step2: Vectorized representation.} To exploit the way the rooted subgraphs compose graphs to learn their embedding, a skip-gram architecture is utilized in the model of Graph2vec to train the distributed representation of subgraphs. The objective of the skip-gram model is to maximize the following log probability:
\begin{equation}
    \sum ^{l_{i}}_{j=1}\log P_{r}\left( sg_{n}|g_{i}\right) 
\end{equation}

where $sg_n$ is the rooted subgraph of n in $G_i$ generated through the above step, $P_r$ denotes the probability that $sg_n$ appears in $g_i$,\revise{ $l_i$ is the total number of rooted subgraphs extracted from graph $g_i$ and $j$ indexes over these subgraphs.} The skipgram model captures shared similarities between rooted subgraphs, so that graphs consisting of similar rooted subgraphs are represented close to each other. Finally, we get a $N_s$-dimensional structure vector $S_i$=$\left\{ S_{i1},S_{i2},\cdots ,S_{iN_s} \right\} $, which represents the global structure of graph $G_i$.

\subsection{Attribute-Structure synchronization}

In this section, we will introduce how to extract the attribute features from graphs and combine them with the structural features. 

1) Attribute feature extraction: The attributes of graphs are extracted based on the domain knowledge of graphs, including micro-level and macro-level. The macro-level attributes refer to global characteristics of graphs, such as the total number of papers published by the authors in a co-authored network, the timespan of a family obtained by subtracting the birth year of the first ancestor from the death year of the last offspring in the genealogical trees, etc., which reflect the graph-oriented information. In contrast, the micro-level attributes can be obtained by directly summing or averaging the values associated with nodes, such as the average age of users in a social community, and the average working hours of members in a working team. 

From the micro-level and macro-level attributes, users can choose appropriate attributes to solve different analysis tasks. For example, to identify teams with strong scientific research ability in a co-author dataset, a variety of attributes can be considered, such as the total number and total cites of papers published by team members, the average author ranking of team members, and the cooperative relationship between team members. When studying the correlation between research productivity and age composition for research teams, the number of published papers and the average age of team members need to be focused on. Finally, these macro-level and micro-level attributes constitute an attribute vector of each graph, \revise{$A_i$=$\left\{ a_{i1}, a_{i2},\cdots, a_{iN_A} \right\} $}, where $a_{ij}$ refers to the value of attribute $j$, $N_A$ refers to the number of attributes which are extracted from the graph $G_i$. 

2) CCA-based Attribute-Structure Synchronization: \linebreak Canonical Correlation Analysis (CCA) \cite{sargin2007audiovisual}, coupling dimension reduction and feature fusion together, can effectively fuse two different features by maximizing their canonical correlation, which is widely used for image and video retrieval, indexing and annotation. Considering its promising performance in classification and identification, we use CCA to perform joint dimensionality reduction by preserving the correlation between attribute features and structure features. 

\revise{
CCA seeks linear projections of the structural and attribute feature spaces such that the resulting canonical variables are maximally correlated. The latent shared components identified by CCA correspond to underlying factors that simultaneously influence both the graph topology and its attribute distribution. For example, a latent component may capture a functional cluster (e.g., a research topic or organizational unit) that manifests as both a tightly connected subgraph (structure) and a set of recurring attribute patterns (e.g., author affiliation, node roles).}

Based on the feature extractions as discussed above, suppose we have obtained the structure feature vectors $\left\{ S_{1}, S_{2}, S_{3},\cdots, S_{M}, \right\} $ and the attribute feature vectors $\{ A_{1}, \allowbreak A_{2}, \dots, A_{M} \}$ of all graphs obtained by the previous two sections in a large graph dataset, where $M$ is the number of graphs. Let $S$ and $A$ be two multidimensional variables in the two sets of feature vectors, and $N_S$ and $N_A$ are their dimensions respectively. Then we normalize different feature value ranges or scales of $S$ and $A$ so that the resulting features have zero mean and unit variance. CCA aims to find two linear transformation matrices, $H_S$ and $H_A$, respectively for $S$ and $A$, to maximize the mutual information between the transformed vectors \revise{ $S'$ and $A'$}.
\begin{equation}
    S'=H_{s}S,A'=H_{A}A 
\end{equation}

\revise{We refer to the pair( $S'$, $A'$ ) as the CCA transform of $S$ and $A$. The dimension of the transformation matrix $H_S$ and $H_A$ are $m  \times N_S$ and $m  \times N_A$ respectively ($m \leq $ min($N_S$,$N_A$)) ,which are represented as follows:}
\begin{equation}
    \revise{
    H_{S}=\left[\begin{array}{c}h_{S 1}^{T} \\ h_{S 2}^{T} \\ \vdots \\ h_{S m}^{T}\end{array}\right], H_{A}=\left[\begin{array}{c}h_{A 1}^{T} \\ h_{A 2}^{T} \\ \vdots \\ h_{A m}^{T}\end{array}\right]
    }
\end{equation}

\revise{The rows of these two matrices, $\left\{h_{S i}\right\}$ and $\left\{h_{A i}\right\}(1\leq i \leq m)$, are considered as CCA basis vectors. The first pair of these basis vectors, $\left\{h_{A i}\right\}(1\leq i \leq m)$, are obtained by the following formula :}
\begin{equation}
\left(h_{S1},h_{A1}\right)=\arg\operatorname{max}_{\left(h_{S},h_{A}\right)}C o r r\left(h_{S}^{T}S,h_{A}^{T}A\right)
\end{equation}

That means the projections of \textit{S} and \textit{A} along the directions represented by $\left(h_{S1},h_{A1}\right)$ are maximally correlated. The first pair of canonical components can be computed by $S'_1$ = $h_{S1}^{T}S$, $A'_1$ = $h_{A1}^{T}A$. Next, the second pair of CCA basis vectors can be extracted by maximizing the same correlation on the basis of ensuring that they are not correlated to the first pair of canonical components. Thus the remaining canonical pairs can be extracted by the same procedure. The basis vectors are often computed by solving the following equivalent eigenvalue problem relating to the joint covariance matrix C, under the constraints of $h_{S}^{T}h_s$ = 1 and $h_{A}^{T}h_A$ = 1.
\begin{equation}
    C=\begin{bmatrix}
  C_{SS} ,&C_{SA}  \\
  C_{AS} ,&C_{AA} 
\end{bmatrix}
\end{equation}
\revise{where $C_{SS}$ and $C_{AA}$ are the within-set covariance matrices, $C_{SA}$ or $C_{AS}$ is the the between-set covariance matrix.} Furthermore, we obtain the eigenvectors, namely the normalized CCA basis vectors, and each associated eigenvalue $\gamma_i$, i = 1,2,$\cdots$, N, represents the canonical correlation. 

Finally, for each graph in the large graph dataset, the original attribute features and structure features can be transformed by CCA to a pair of new vectors with the same length, which is then directly concatenated as an attribute-structure fused vector.

\begin{figure*}[t]%
\centering
\includegraphics[width=1\textwidth]{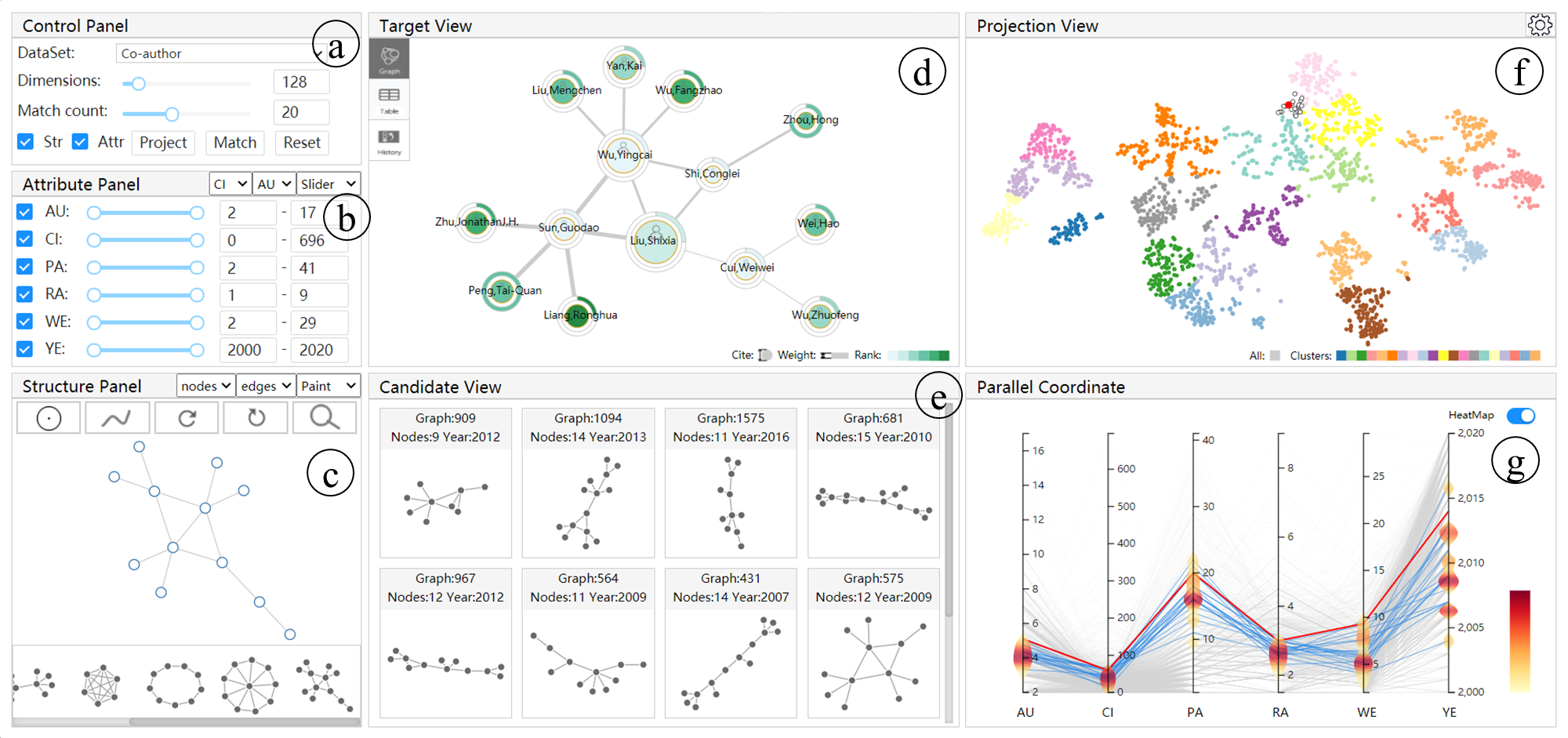}
\caption{A case study is conducted to retrieve research communities with similar collaboration and attributes from a large amount of co-author network datasets. (a) The control panel enables users to specify the parameters of our graph matching model. (b) The attribute panel enables users to filter attributes and specify their value ranges. (c) In the structure panel, users can specify desired structures as the target graphs. (d) The matched graph of interest is highlighted in the target view with attributes encoded with different visual elements. (e) The candidate view presents the other matched graphs in the form of a structure overview. (f) Graph2vec is conducted to combine the representations of structures and attributes, and transform graphs into a projection view in which graphs located nearby share similar structures and attributes. (g) The parallel coordinate view is designed to present detailed attributes of the matched graphs.}\label{fig1}
\end{figure*}

\subsection{Graph matching}

Based on the attribute-structure synchronization using CCA, all graphs are represented as high-dimensional fused vectors by synthesizing structures and attributes. \revise{To support exploratory analysis, we conduct clustering on the fused vectors to help users gain an overview of the embedding space. By observing representative graphs within different clusters, users can develop an initial understanding of how structural and attribute features are distributed across the dataset, which facilitates informed exploration and comparison.} The density-based clustering algorithm DBSCAN \cite{ester1996density} and the distance-based clustering algorithm $K$-means \cite{macqueen1967some} are both used here. Thus, users can decide which clustering method to use and adjust their parameters according to the distribution characteristics of graphs in the projection space. When a graph is specified as the target graph, all graphs within the cluster this target graph belongs to are regarded as the matching results. Furthermore, an alternative graph matching is based on the $K$ -nearest neighbor ($k$-NN), which takes the first $K$ (users can specify the number of matching graphs) graphs that are closest to the target graph as the matching results in the original high-dimensional vector space obtained by CCA.

\section{Visual Interfaces}
\label{sec5}

We develop a visual graph matching system that consists of multiple coordinated views (target view, candidate view, projection view, and parallel coordinate view), enabling users to perform graph matching and evaluate the results from both structural and attribute perspectives in an exploratory manner.

\subsection{Interactive Graph Matching}
\label{sec5.1}

Users can specify the target graph through two modes: navigation \& clustering-based graph matching, and feature-definition-based graph matching.

\subsubsection{Navigation \& Clustering-Based Graph Matching}
\revise{
To support users in exploring the entire dataset and identifying graphs of interest, we provide two overview visualizations—the 2D attribute scatter-plot view and the projection view—along with a control panel that allows users to configure matching and visualization parameters.
}

\textbf{Control Panel.}
\revise{
The control panel (Figure~\ref{fig1}(a)) enables users to configure key parameters that influence the graph matching process. It allows switching among three modes: \emph{structure-based}, \emph{attribute-based}, or a \emph{combined structure-attribute} fusion. If both are selected, the system activates a CCA-based model to embed all graphs into a shared latent space.
}

\textbf{Attribute Scatter-Plot View.}
\revise{
The 2D attribute scatter-plot view presents graphs as data points based on user-specified attributes along the x- and y-axes (Figure~\ref{fig5}(a)). This helps users understand the overall attribute distribution and inter-attribute relationships. Both the attribute and structure panels offer this view: the former visualizes semantic attributes, while the latter shows structural metrics (e.g., number of nodes).
}

\textbf{Projection View.}
\revise{
The projection view (Figure~\ref{fig1}(f)) offers a two-dimensional summary of the dataset using t-SNE~\cite{cai2022theoretical}, where the distance between scatter points reflects graph similarity. Users can toggle among structural, attribute, or fused embeddings as configured in the control panel. The system supports clustering via DBSCAN or $k$-means, and cluster memberships are color-coded to reveal graph groups and outliers.
}

\subsubsection{Feature-Definition-Based Graph Matching}
\revise{
Users may also construct a target graph manually by defining its structure and attribute features. Attributes are adjusted using sliders in the attribute panel (Figure~\ref{fig1}(b)). Structural exemplars can be drawn manually or selected from common patterns (templates) extracted from frequent Graph2vec clusters (Figure~\ref{fig1}(c)). For these newly created graphs, Graph2vec is applied online to compute their embeddings for matching.
}

\subsection{Graph Matching Evaluation}

\textbf{Target View.}
\revise{
When a target graph is selected—from either the projection view or through custom creation—it appears in the target view (Figure~\ref{fig1}(d)). Two layout options are provided: a graph layout and a tree layout. The tree layout highlights temporal or hierarchical structures along a time axis. Nodes and edges are encoded with color, size, and shape to reflect attributes and structure. Legends assist interpretation, and interactions like zooming and panning support in-depth analysis.
}

\textbf{Candidate View.}
\revise{
The candidate view (Figure~\ref{fig1}(e)) shows all matched graphs using node-link diagrams. Candidates are ranked by similarity and the number displayed is adjustable. Selecting a candidate updates the target view for side-by-side comparison.
}

\textbf{Parallel Coordinates View.}
\revise{
To compare attribute similarity, the parallel coordinates view (Figure~\ref{fig1}(g)) plots graphs as polylines across multiple attribute dimensions. A density stream overlays each axis, encoding the distribution of values by color and width. When matches are identified, their lines are highlighted, and the streams update to reflect the attribute alignment with the target.
}

\section{Evaluation}
\label{sec6}

Our visual graph matching system is developed by a classic web-based framework "flask+d3.js+python+MongoDB". Quantitative comparisons of two datasets are conducted to evaluate the effectiveness of our matching method. Case studies and expert feedback are also detailed in this section.

\subsection{Datasets}
Two real-world attributed graph datasets are applied to conduct case studies. 

\textbf{Genealogy dataset:} The China Multigenerational Panel Dataset-Liaoning (CMGPD-LN) \cite{lee2010china} is transcribed from the population registers compiled by the Qing Dynasty government in Liaoning Province, northeastern China, from 1749 to 1909. This dataset, with more than 1.5 million records, provides socioeconomic, demographic, and other characteristics for over 260,000 individuals. According to the relationship between individuals, we build approximately 12, 000 family trees. In addition, five attributes are extracted and constructed from member information to characterize the features of families, which include the timespan of a whole family (TS), the average age of family members (AA), the number of family members who have been officials (PN), the number of villages where family members have been lived (VN), and the average gap between father and son (AG). 

\textbf{Co-author network dataset:} The papers published from 2001 to 2020 in three research areas of visualization (TVCG), data mining (KDD), and human-computer interaction (CHI) are crawled from \href{https://www.webofscience.com/}{Web of Science}. For each year, we construct a large co-author network according to the co-authorship between scholars of the year. Then the module-based community detection method \cite{li2021community} is applied to all the networks, and about 4,000 small co-author networks, namely research communities, are obtained. We construct multiple attributes for each research community, such as the average number of authors per paper (AU), the average citation count per paper (CI), the total number of published papers (PA), the publication year (YE), the average author rank (RA), and the average number of co-authored papers between a pair of team members (WE).

\subsection{Quantitative comparison}

We present our experimental studies for the proposed graph matching method based on CCA, which supports attributed graph matching through attribute-structure synchronization. We compare our method with four matching strategies. \textbf{(1)Structures (Str)}  measures graph similarity by the graph embedding technology, Graph2vec, without considering graph attributes. \textbf{(2) Attributes (Attr)} measures graph similarity based on multi-dimensional attributes, without considering graph structures. \textbf{(3) Direct concatenating (DC)} directly concatenates the high-dimensional structure vectors (learned from Graph2vec) and low-dimensional attribute vectors, then jointly reduces the dimensions to measure the graph similarity. \textbf{(4) Indirect concatenating (IDC)} reduces two different feature vectors into a unified dimension and concatenates them for measuring the graph similarity.


For each dataset, we first randomly select twenty graphs as the target graphs and apply the $K$-nearest neighbor to find their matching results in the embedding space generated by the above matching methods. To further compare the effectiveness of different graph matching methods, we respectively evaluate the quality of matching results from aspects of structure and attribute. (1) Structure similarity. As a classic structure similarity measure, graph editing distance (GED) \revise{\cite{moscatelli2024graph}} refers to the minimum operation cost (addition, deletion, and substitution) of transforming from one graph to another. We compute the average GED between each target graph and its matched graphs as the evaluation metric in terms of structure similarity. (2) Attribute similarity. Euclidean distance is a widely used method to measure the similarity between multi-dimensional vectors \revise{\cite{wills2020metrics}}. The average Euclidean distance of the attribute vectors between each target graph and its matched graphs is considered as the evaluation metric in terms of attribute similarity. 

Table 1 summarizes the experimental results across different datasets under various $K$ values ( $K$ -nearest neighbor). Since the number of genealogies is relatively large, we set larger $K$ values for it. Through the comparison results of the genealogy dataset, it can be found that our method performs better than Str, Attr, DC, and IDC in terms of structure similarity. When $K$ is specified as 20, our method performs a little inferior to Str and DC. With the increase of $K$, our method outperforms the other methods. In terms of attribute similarity, our method performs inferior to Attr and performs better than Str, DC, and IDC. This indicates that our method combines the structure and attribute of graphs to strike the balance between structure similarity and attribute similarity. In addition, our method almost performs better than DC and IDC in terms of structure similarity and attribute similarity, which proves that our method effectively overcomes the shortcomings of conventional fusion methods. For the coauthor network dataset, the experimental results reach the same conclusion. Besides, our method goes beyond Attr in terms of attribute similarity unexpectedly, which demonstrates that CCA retains the maximization of mutual information between structures and attributes. In addition to CCA, we conducted experiments with the Kernel Canonical Correlation Analysis (KCCA). The experimental results, available in our supplementary material, indicate that KCCA did not perform well. This ineffectiveness may be attributed to KCCA's mapping of low-dimensional data into a high-dimensional kernel function space, resulting in the generation of redundant features between the data while simultaneously increasing computational overhead. Furthermore, the intervention of nonlinear kernel functions has the effect of perturbing the linear relationship between structures and attributes, potentially resulting in the loss of certain features. The above results demonstrate that our CCA-based matching effectively maintains the similarity of structures and attributes.

\begin{table}[]
\centering
\resizebox{\linewidth}{!}{
\begin{tabular}{cccccccc}
\hline
Dataset                                                                                & k                   & Type     & Str           & Attr           & Our            & DC            & IDC   \\ \hline
\multirow{6}{*}{\begin{tabular}[c]{@{}c@{}}Genealogy\\ Dataset\end{tabular}}           & \multirow{2}{*}{20} & Str-Sim  & \textbf{9.64} & 10.59          & 10.01          & 9.74          & 40.06 \\
                                                                                       &                     & Attr-Sim & 63.08         & \textbf{7.02}  & 12.53          & 20.89         & 21.66 \\
                                                                                       & \multirow{2}{*}{40} & Str-Sim  & 9.29          & 9.43           & \textbf{8.95}  & 9.01          & 9.11  \\
                                                                                       &                     & Attr-Sim & 61.83         & \textbf{7.97}  & 13.04          & 20.97         & 22.15 \\
                                                                                       & \multirow{2}{*}{60} & Str-Sim  & 7.97          & 7.86           & \textbf{7.45}  & 7.79          & 7.92  \\
                                                                                       &                     & Attr-Sim & 63.69         & \textbf{10.20} & 15.36          & 22.49         & 24.17 \\ \hline
\multirow{6}{*}{\begin{tabular}[c]{@{}c@{}}Co-author\\ Network\\ Dataset\end{tabular}} & \multirow{2}{*}{5}  & Str-Sim  & \textbf{5.81} & 6.04           & 5.92           & 5.90          & 5.99  \\
                                                                                       &                     & Attr-Sim & 35.61         & 13.52          & \textbf{9.43}  & 15.39         & 20.47 \\
                                                                                       & \multirow{2}{*}{10} & Str-Sim  & 6.16          & 6.17           & 6.14           & \textbf{6.12} & 6.24  \\
                                                                                       &                     & Attr-Sim & 36.19         & 17.85          & \textbf{11.60} & 19.90         & 23.15 \\
                                                                                       & \multirow{2}{*}{15} & Str-Sim  & 5.99          & 5.92           & \textbf{5.90}  & 5.97          & 6.01  \\
                                                                                       &                     & Attr-Sim & 38.39         & 20.38          & \textbf{12.91} & 22.05         & 25.57 \\ \hline
\end{tabular}}
\caption{Quantitative Comparison}

\end{table}

\revise{
To further evaluate the robustness of our CCA-based framework under varying neighborhood sizes, we conducted additional experiments on the Co-author dataset using different values of $k \in \{5, 10, 15\}$. As shown in Table 2, our method consistently yields low structure and attribute matching errors across different k values, outperforming KCCA, DC, and IDC. These results indicate that our framework remains effective under varying levels of local graph context, further validating its generalizability.
}
\begin{table}[htbp]
\centering
\begin{tabular}{c c c c c c c c}
\toprule
\multicolumn{1}{c}{\textbf{k}} & \textbf{Type} & \textbf{Str} & \textbf{Attr} & \textbf{Our} & \textbf{KCCA} & \textbf{DC} & \textbf{IDC} \\
\midrule
\multirow{2}{*}{5} & Str & \textbf{5.81} & 6.04 & 5.92 & 6.79 & 5.90 & 5.99 \\
 & Attr & 35.61 & 13.52 & \textbf{9.43} & 20.72 & 15.39 & 20.47 \\
\cmidrule{1-8}
\multirow{2}{*}{10} & Str & 6.16 & 6.17 & 6.14 & \textbf{6.10} & 6.12 & 6.24 \\
 & Attr & 36.19 & 17.85 & \textbf{11.60} & 22.16 & 19.90 & 23.15 \\
\cmidrule{1-8}
\multirow{2}{*}{15} & Str & 5.99 & 5.92 & \textbf{5.90} & 5.93 & 5.97 & 6.01 \\
 & Attr & 38.39 & 20.38 & \textbf{12.91} & 22.91 & 22.05 & 25.57 \\
\bottomrule
\end{tabular}
\caption{Robustness Evaluation of CCA}
\label{tab:quant-compare}
\end{table}

\subsection{Case studies}

Given the above two real-world attribute graph datasets, case studies are conducted to demonstrate the effectiveness of our attribute-structure synchronization graph matching system.

\begin{figure}[h]
    \centering
    \includegraphics[width=\linewidth]{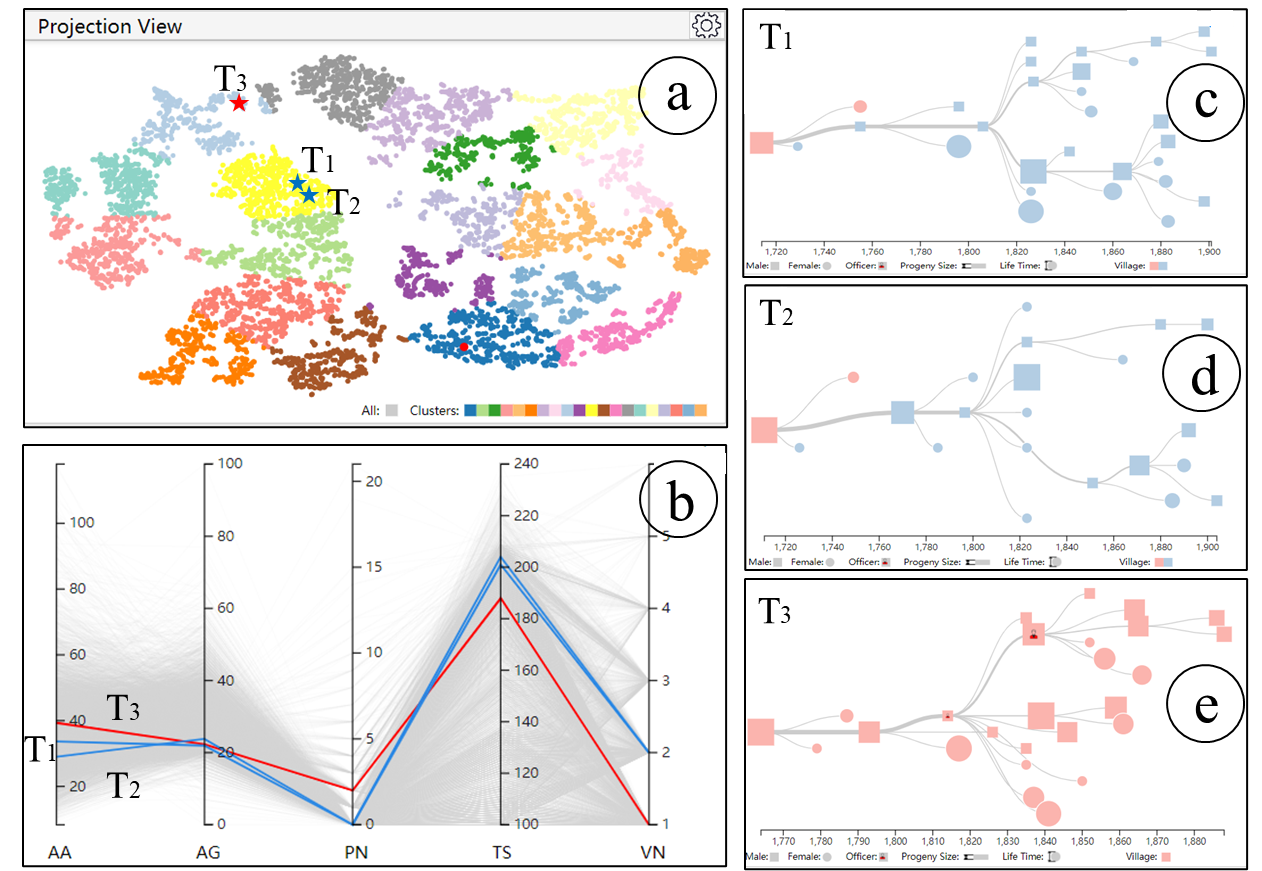}
    \caption{Exploration on the projection view (a) with structure features and attribute features considered together. (c), (d), and (e) are family trees with similar structures, but (e) is apart from the other two due to the attribute difference, which is revealed by the parallel coordinate view (b).}
    \label{fig4}
\end{figure}

1) Genealogy matching: We invited $E_3$, a social science expert, to explore the genealogy dataset (D1) using our system. He loaded the dataset via the control panel and examined the family distribution in the projection view (Figure~\ref{fig4}(a)). By clicking several points, he noticed the structural complexity of family trees decreased from left to right. He commented, “\emph{The distribution of structures shows that the attribute-structure synthetical space preserves the structure similarity well on the whole.}”

\begin{figure*}[h]
    \centering
    \includegraphics[width=\linewidth]{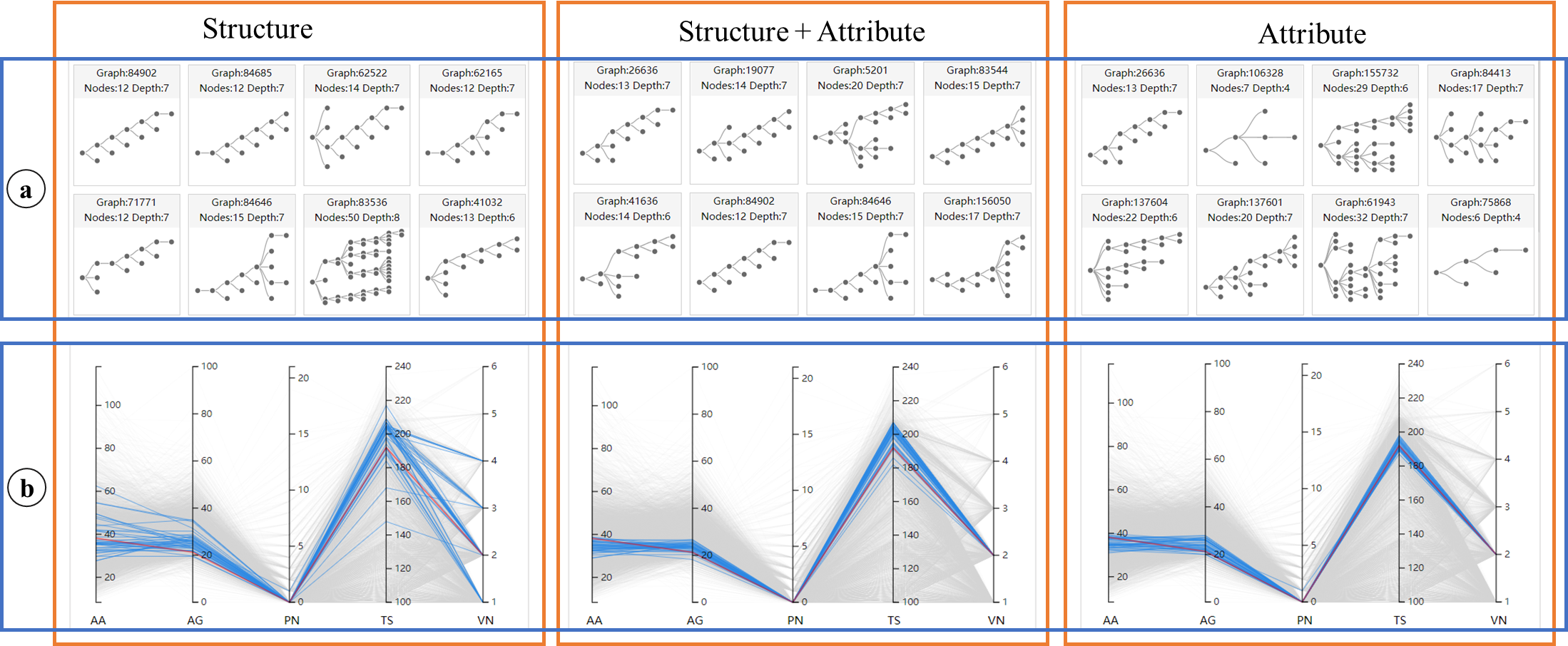}
    \caption{Given the target graph in Figure ~\ref{fig5}(b), the matching results under different similarity measures are generated, which are evaluated in terms of structure similarity through the candidate view (a), and attribute similarity through the parallel coordinate view (b).}
    \label{fig6}
\end{figure*}
Next, $E_3$ explored a cluster and found nearby points (e.g., T1 and T2 in Figure~\ref{fig4}(c, d)) had similar structures and attributes. The structure and attribute information associated with nodes of the selected graph were displayed in detail in the target view. For each node, its size and shape respectively encode the lifespan and gender (a circle represents a female, and a square represents a male). Besides, the color encodes the village where he/she lived. For each edge, its width encodes the number of offspring. In addition, if someone has ever been an official, there will be a mark on the corresponding node. $E_3$ said, ``\emph{Structure is an abstract term for me, but the target view based on the node-link graph provides a straightforward way to understand the structure of the family tree.}" He added, ``\emph{Attribute mappings give a deeper understanding of family characteristics.}" After he explored other points in adjacent clusters, he found that there was a family tree T3 as shown in Figure ~\ref{fig4}(e), of which the structure was almost the same with T1 and T2. Then he carefully examined the parallel coordinate view as shown in Figure ~\ref{fig4}(b), to compare the attributes of T1, T2 and T3. $E_3$ found that though T3 was similar to T1 and T2 in terms of structure, its attribute information was different from them, especially on the attributes of ``PN", ``TS" and ``VN". $E_3$ said, ``\emph{This finding is an excellent demonstration of the effectiveness of the CCA-based method, which takes structures and attributes into consideration synchronously.}"

\begin{figure}[h]
    \centering
    \includegraphics[width=\linewidth]{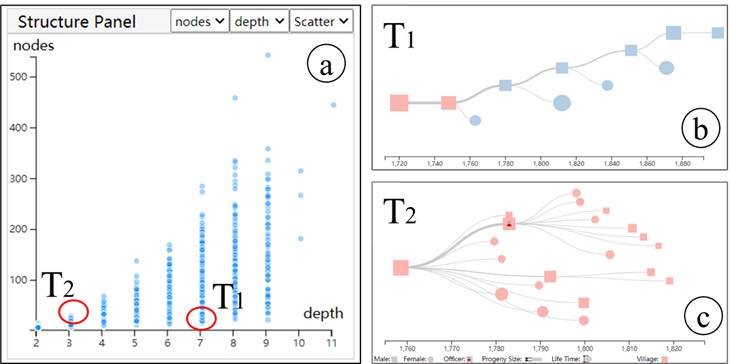}
    \caption{Two graphs with special features are selected from the 2D attribute scatter-plot view. (a) The x-axis represents the depth (generations) of the family tree, and the y-axis represents the number of nodes in the family tree. (b) is a single-linage family with a small number of nodes and a large depth. (c) is an inclined family, with a relatively large number of nodes and a small depth.}
    \label{fig5}
\end{figure}

Then, the 2D attribute scatter-plot view in the structure panel caught the attention of $E_3$. He set the x-axis to the ``depth" representing the number of generations, and the y-axis to the ``nodes" representing the total number of family members, as shown in Figure ~\ref{fig5}(a). When the expert clicked a point T2 in the region with a number of nodes and a small depth (Figure ~\ref{fig5}(a)), he found this was a typical inclined family. As shown in Figure ~\ref{fig5}(c), the eldest family member of the second generation had many descendants, and he also had the experience of being an official. ``\emph{This is very much in line with the Chinese tradition of primogeniture.}" the expert commented. Then the expert clicked another point T1, a single-lineage family with many generations but few members, as shown in Figure ~\ref{fig5}(b). ``\emph{This overview makes exploring distinctive families much easier.}" $E_2$ praised.

$E_3$ turned his attention to the matching function and selected the above single-linage family (Figure ~\ref{fig5}(b)) as a target graph. Through the control panel, $E_3$ set three different similarity measures, namely structure, attribute, and ``structure + attribute". Matching results were visualized in Figure~\ref{fig6}, with structures shown in the candidate view (Figure~\ref{fig6}(a)) and attributes in the parallel coordinate view (Figure~\ref{fig6}(b)). After examining the candidate views (Figure ~\ref{fig6}(a)), he found the structure similarities of the matching results obtained by our method were retained well compared with ``Str" and ``Attr". Then he examined the parallel coordinate view as shown in Figure ~\ref{fig6}(b), and found ``Str" had a very random distribution of attribute values. However, our method was indistinguishable from ``Attr" on the attribute distribution of the matching results. Specifically, both of them performed equally well on ``AA", ``PN" and ``VN", but ``Attr" outperformed our method on ``TS", and was a little worse than our method on ``AG". $E_2$ thought, ``The CCA-based method performs better than traditionary structure or attribute-based matching methods in both structure similarity and attribute homogeneity of resultant matching graphs."

\begin{figure}[h]
    \centering
    \includegraphics[width=\linewidth]{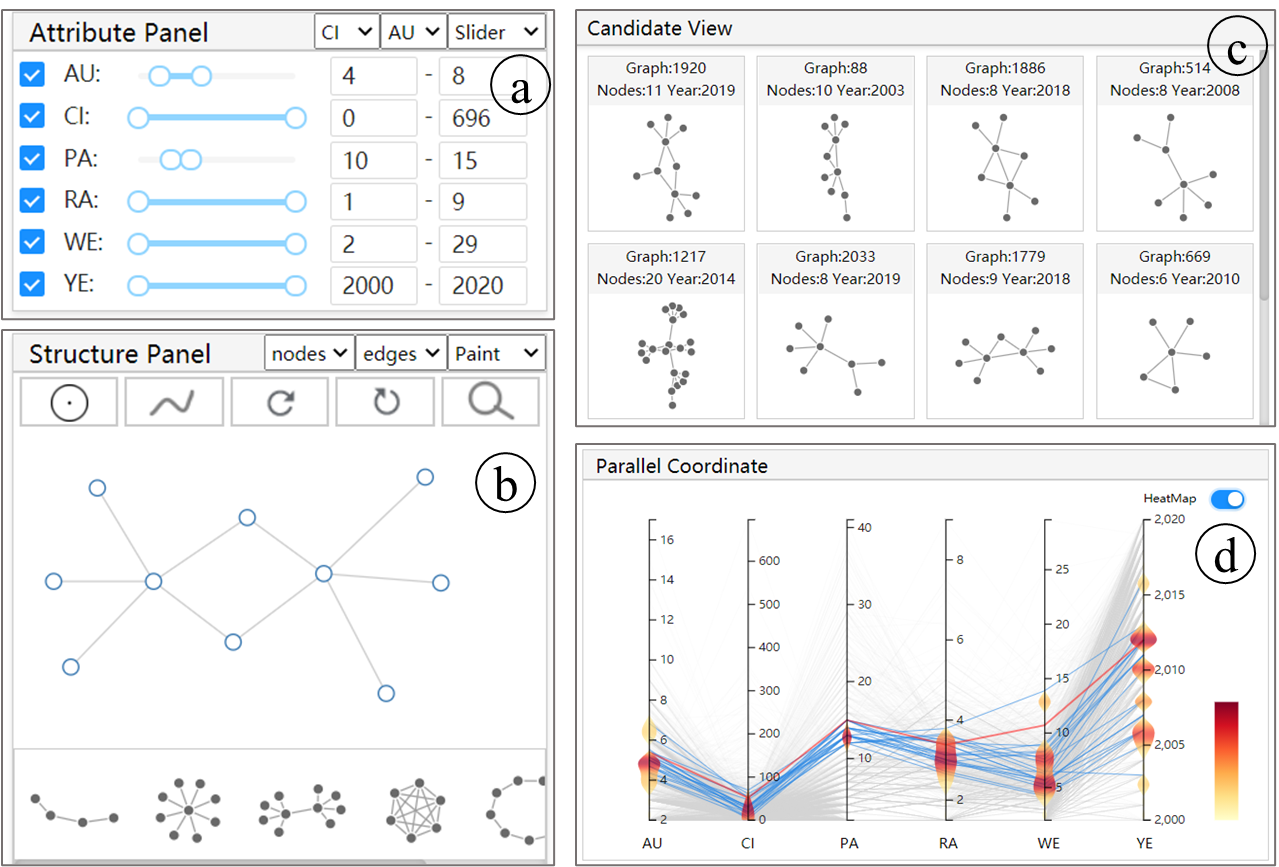}
    \caption{Definition and matching of a target graph. (a) and (b) enable users to specify the attribute and structure of the target graph. (c) and (d) show the detailed information of matching results.}
    \label{fig7}
\end{figure}

\begin{figure}[h]
    \centering
    \includegraphics[width=\linewidth]{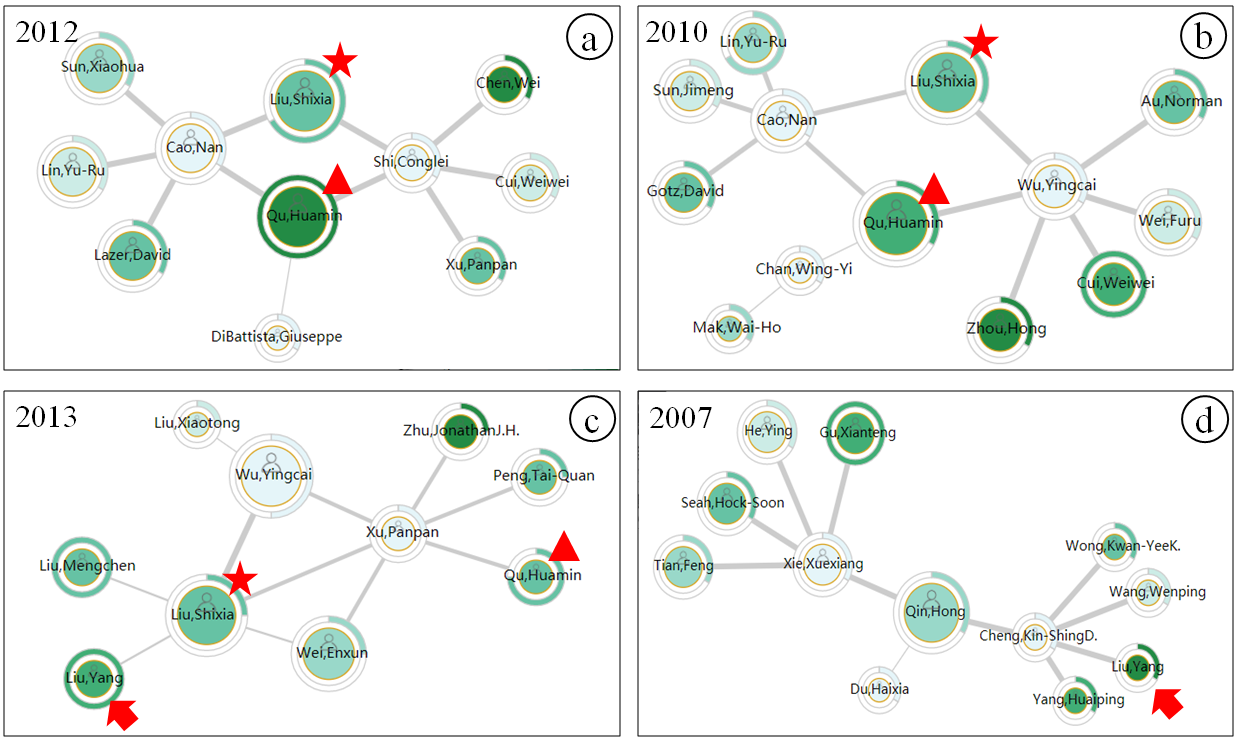}
    \caption{Comparison of four matching results in Figure 7(c).}
    \label{fig8}
\end{figure}

2) Co-author network exploration: Experts $E_1$ and $E_2$ explored the co-author network dataset D2. As members of two research groups with collaboration ties, they aimed to find similar cooperation patterns. They sketched a structure with two core nodes connecting star-shaped groups (Figure~\ref{fig7}(b)). They adjusted the value ranges on attributes through the attribute panel, as shown in Figure ~\ref{fig7}(a). For example, the average number of authors (AU) listed in each paper was set between 4 and 8. The total number of papers published by the group (PA) was set between 10 and 15. After the structures and attributes were defined, $E_1$ clicked the buttons ``Project" and ``Match" in turn. The matching results were displayed in the candidate view, as shown in Figure ~\ref{fig7}(c). Among them, the best one was shown in the target view as shown in Figure ~\ref{fig8}(a). In the target view: node size encodes citation count; outer ring fill ratio shows paper count; node color reflects average author rank (darker = lower rank); and edge width indicates collaboration frequency.

$E_2$ noticed a node labeled “Huamin Qu” (red triangle) with the darkest color and largest radius in Figure~\ref{fig8}(a). He speculated ``\emph{this author may be a corresponding author with many cited papers and broad collaborations.}". $E_1$ then explored other graphs and found that Shixia Liu, Huamin Qu, Nan Cao, and Yingcai Wu appeared frequently. Shixia Liu was marked as a core author (red star) in multiple graphs (Figure~\ref{fig8}(a, b, c)), and Yingcai Wu and Huamin Qu often appeared as core authors together. $E_2$ noted that ``\emph{ senior scholars often maintain long-term collaborations.}". They also found that ``\emph{Liu’s co-author networks from 2010–2013 matched, but not in 2014. By querying her name, they examined her 2014 co-author network in Figure~\ref{fig1}(d), discovering her team expanded from two to three groups.}".

After exploring the matching results carefully, $E_1$ and $E_2$ paid attention to Yang Liu (highlighted by a red arrow) who appeared in both the cooperation networks of 2007 and 2013. They were surprised to find that though the two cooperation networks were very similar in terms of structure and attributes, other members of the two networks didn’t overlap, as shown in Figure ~\ref{fig8}(c) and Figure ~\ref{fig8}(d). Then $E_2$ clicked the ``Table" button in the target view to check his published papers. He found that Yang Liu focused on geometric modeling and processing in 2007, and he collaborated with experts in this field. In 2013, he turned his attention to visualization and closed cooperation with experts in the field of visualization. Finally, the experts concluded that our matching model was conductive to learn the correlation between structures and attributes, and improved the accuracy of matching results in terms of both structure similarity and attribute similarity.

\revise{3) In addition to visual and quantitative validation, we further evaluated the runtime performance of our framework and baseline methods. The runtime comparison was conducted on both the Genealogy and Co-author datasets with different values of k, using a workstation with an AMD R7 CPU and 32GB of RAM.
As shown in Table 3, our method achieves comparable efficiency to traditional approaches while delivering superior graph matching quality. This demonstrates that the proposed framework is not only accurate and interpretable but also computationally feasible for practical usage.}
\begin{table}[htbp]
\centering
\label{tab:time-consumption}
\begin{tabular}{@{}l c c c c c c@{}}
\toprule
\textbf{Dataset} & \textbf{k} & \textbf{Str} & \textbf{Attr} & \textbf{Our} & \textbf{DC} & \textbf{IDC} \\
\midrule
\multirow{3}{*}{Genealogy} 
 & 20 & 75.085 & 73.812 & 79.087 & 72.112 & 72.661 \\
 & 40 & 75.047 & 75.857 & 79.010 & 72.328 & 73.785 \\
 & 60 & 75.147 & 74.011 & 79.268 & 72.594 & 74.339 \\
\cmidrule{1-7}  
\multirow{3}{*}{Co-Author}
 & 5 & 13.356 & 14.722 & 15.040 & 13.438 & 13.669 \\
 & 10 & 13.449 & 14.894 & 15.041 & 13.561 & 14.493 \\
 & 15 & 13.331 & 14.915 & 15.041 & 14.602 & 14.984 \\
\bottomrule
\end{tabular}
\caption{System Time Consumption}

\end{table}

\section{Disccusion}
\label{sec7}
\revise{
In this section, we summarize the limitations of our modular graph matching framework and discuss its generalizability and the scope of the visual analytics system.}
\revise{
\subsection{Limitations of the Framework}}

\revise{
The proposed graph matching framework adopts graph2vec for structural embedding and Canonical Correlation Analysis (CCA) for aligning structural and attribute representations. This design reflects a decoupled and modular approach to integrating heterogeneous graph information, in which each component can be independently configured or substituted. For experimental validation, we selected baseline strategies such as direct and indirect concatenation to emphasize the contribution of the proposed alignment mechanism.
}

\revise{
While effective in demonstrating the feasibility of modular structure-attribute integration, this design presents several limitations. First, graph2vec encodes graph-level structure into a fixed embedding space, which may not capture finer-grained topological semantics or hierarchical substructures as effectively as more expressive neural encoders. Second, the use of CCA provides a computationally efficient and analytically tractable means for cross-view alignment, but is limited in its capacity to model non-linear or high-order dependencies between structural and attribute spaces.
}

\revise{Recent developments in graph neural networks (GNNs) offer a powerful end-to-end paradigm for integrating structure and attributes through message passing. While potentially yielding better performance, GNN-based models often act as black boxes, making it difficult to interpret or intervene in the matching process. In contrast, our framework adopts a modular design that separates structure encoding (e.g., via graph2vec) and cross-view alignment (e.g., via CCA), offering greater transparency and flexibility.}

\revise{This design is particularly advantageous in scenarios requiring interpretability and domain knowledge integration, such as financial risk control or medical analysis, where embedding semantics must remain accessible and adjustable. The modular nature also allows each component to be replaced with more advanced alternatives, including GNNs, non-linear alignments, or kernel-based methods, depending on data characteristics or task requirements.}

\revise{Additionally, to further enhance adaptability, a user-controllable weighting mechanism could be introduced to balance structural and attribute importance during matching. Although not implemented in the current version, such extensions can be naturally supported by our flexible architecture.}

\revise{
\subsection{Generalizability and Visual design}}
\revise{
While our framework is designed to be modular and extensible, its application to graph datasets from other domains involves several practical considerations. In particular, structural embeddings can be reused across domains, but attribute representations often vary significantly in terms of dimensionality, semantics, and availability. To accommodate such variability, preprocessing steps such as attribute normalization, feature selection, or domain-specific encoding may be required. The alignment module (e.g., CCA) assumes a common latent space for structure and attributes, so adapting the framework to new domains may involve retraining or substituting the alignment model based on the statistical properties of the new data.
}

\revise{
Regarding the visual analytics component, it is primarily developed to support the evaluation and interpretation of our graph matching method. The interface facilitates interaction with matched graph pairs and highlights attribute-structure correspondence, but it is not designed as a domain-agnostic tool. As such, customization for new datasets may require direct modification of interface parameters or code-level adjustments, particularly when attribute types, graph schemas, or visual encoding conventions differ. Although the current system supports basic parameter configuration (e.g., threshold tuning, similarity metrics), deeper adaptation still relies on developer intervention rather than end-user customization.
}

\revise{
In addition, while the visual interface qualitatively enhances interpretability and supports human-in-the-loop analysis, its effectiveness has not been formally quantified. Future work could explore user studies or task-based evaluations to assess how the visual interaction aids understanding, decision-making, or performance in different matching scenarios.
}

\section{Conclusion}
\label{8}

In this paper, Graph2vec is utilized to represent the structures of graphs, and a Canonical Correlation Analysis model is designed to combine the structure and attribute information into a synthetical embedding space. A distance-based graph matching scheme is employed to quickly retrieve graphs sharing similar structures and attributes. In addition, a set of visual interfaces are provided enabling users to interactively perform graph matching and visually evaluate the similarity of matching results from various perspectives. Quantitative comparisons and case studies based on real-world datasets have demonstrated the effectiveness of our system.

\section{Acknowledgements}
This work was supported in part by the National Natural Science Foundation of China (No.62277013, No.62177040), the National Statistical Science Research Project (No.2023LZ035), China (No.2023C01119), the Zhejiang Provincial Science and Technology Plan Project (2023C01120), the Public Welfare Plan Research Project of Zhejiang Provincial Science and Technology Department (No.LTGG23H260003, No.LGF22F020034), Open Project Program of the State Key Laboratory of CAD\&CG (No.A2301), Zhejiang Provincial Natural Science Foundation of China (No.LTGG24F020006).






\makeatletter
\renewcommand{\thetable}{\arabic{table}}
\makeatother



\bibliographystyle{elsarticle-num} 
\bibliography{document.bib}






\end{document}